# Band alignment of two-dimensional lateral heterostructures

Junfeng Zhang[1,2], Weiyu Xie[3], Jijun Zhao[2*], Shengbai Zhang[3*]

[1] *School of Physics and Information Engineering, Shanxi Normal University, Linfen 041004, China*

[2] *Key Laboratory of Materials Modification by Laser, Ion and Electron Beams (Dalian University of Technology), Ministry of Education, Dalian 116024, China*

[3] *Department of Physics, Applied Physics, and Astronomy, Rensselaer Polytechnic Institute, Troy, NY 12180, USA*

**ABSTRACT**: Recent experimental synthesis of two-dimensional (2D) heterostructures opens a door to new opportunities in tailoring the electronic properties for novel 2D devices. Here, we show that a wide range of lateral 2D heterostructures could have a prominent advantage over the traditional three-dimensional (3D) heterostructures, because their band alignments are insensitive to the interfacial conditions. They should be at the Schottky-Mott limits for semiconductor-metal junctions and at the Anderson limits for semiconductor junctions, respectively. This fundamental difference from the 3D heterostructures is rooted in the fact that, in the asymptotic limit of large distance, the effect of the interfacial dipole vanishes for 2D systems. Due to the slow decay of the dipole field and the dependence on the vacuum thickness, however, studies based on first-principles calculations often failed to reach such a conclusion. Taking graphene/hexagonal-BN and $MoS_2$/$WS_2$ lateral heterostructures as the respective prototypes, we show that the converged junction width can be order of magnitude longer than that for 3D junctions. The present results provide vital guidance to high-quality transport devices wherever a lateral 2D heterostructure is involved.

---

* Email: zhaojj@dlut.edu.cn (J. J. Zhao), zhangs9@rpi.edu (S. B. Zhang).

# 1. Introduction

Two-dimensional (2D) materials are attractive for their unique physical properties. Graphene [1], hexagonal BN (*h*-BN) [2], $MoS_2$ and $WS_2$ [3, 4], are just some of the well-known examples. To combine the advantages of two different 2D materials, electronic or optoelectronic devices made up of stacked or in-plane heterostructures have been widely exploited [5]. A variety of vertical or lateral heterostructures formed by graphene, *h*-BN, phosphorene, and transition metal dichalcogenides, have been successfully fabricated [6-9].

Concerning lateral heterostructures, Ci et al. [10] first synthesized large-scale monolayer *h*-BNC sheets made of *h*-BN and graphene domains. Levendorf et al. [11] developed a "patterned regrowth", which allows for a spatially-controlled synthesis of graphene/*h*-BN (G/BN) heterostructures. Other groups [12-22] have fabricated G/BN heterostructures on Ru, Rh, Cu, Ni, Au, Ir, SiC, and $SiO_2$ substrates. Prototypical 2D devices, such as integrated circuit [11], field-effect transistor [14, 17], and closed-loop resonator [17], have been demonstrated. New physical properties, including tunable band gap [23], robust half-metallicity [24], and unique thermal transport properties [25], have been predicted theoretically.

Another example of 2D lateral heterostructures is the $MoS(Se)_2$/$WS(Se)_2$ semiconductor heterostructures [26-30]. In particular, $MoS_2$/$WS_2$ [27] and $MoS_2$/$WSe_2$ [29] heterostructures with an atomically abrupt interface along armchair (AC) or zigzag (ZZ) directions have been synthesized. Interesting physical properties for device applications, such as the rectification and photovoltaic effects, have been demonstrated experimentally [26, 27, 29].

Concerning vertical heterostructures, an entire class of them are made of by placing one atomically-thin 2D material (A) on top of another atomically-thin 2D material (B) [31-35], see

Figure 1a for a schematic plot. Due to the thinness of the stacked region, a classical division of the material into a vertical junction between A and B may not be meaningful. Rather, the stacked region should be treated as a new quantum system AB as a whole. Thus, such a vertical heterostructure is not different from having two lateral junctions in series between A and AB and between AB and B. A special case of the stacked heterostructure would be A on A, which makes a lateral 2D junction between A and AA (see Figure 1b), as recently demonstrated experimentally for $MoS_2$ [36] and $MoSe_2$ [37]. Even within a single-phase 2D material, one may effectively set up a lateral junction if one part of the material has been heavily doped, alloyed, or merely being placed near a dielectric media.

While paving the way for novel electronics, the experimental successes call for a fundamental understanding of the electronic properties of these 2D in-plane metal/semiconductor (M/S) and semiconductor/semiconductor (S/S) heterostructures. The most important parameter that determines the transport across the interface is [38, 39], in the case of M/S, the Schottky barrier height (SBH), which is the mismatch between the band edges of the semiconductor and the Fermi level of the metal, or the band offset in the case of S/S, which is the mismatch between the band edges of the two semiconductors. Although the SBHs and band offsets for three-dimensional (3D) heterostructures have been studied extensively in the past, it is unclear whether the traditional wisdom in 3D systems is still applicable to 2D heterostructures. For example, the interfacial dipole is known to play a crucial role in determining the SBHs and band offsets for the 3D M/S and M/M heterostructures, respectively [39, 40]; in 2D heterostructures, on the other hand, a classical electrostatic consideration suggests that the dipole potential should vanish, instead.

In this paper, we address the fundamental difference between 2D and 3D junctions by first-principles calculations and show that the results do approach the limits set by the electrostatic dipole-line model. In other words, due to the decay of interfacial dipole potential in the asymptotic limit of large distance, band alignments for 2D lateral junctions with sufficient width are insensitive to interfacial conditions. Rather, they are determined purely by intrinsic properties of the component materials, namely, the universal Schottky-Mott limits for M/S interfaces and Anderson limits for S/S interfaces, respectively. This is in startle contrast to 3D heterostructures where such limits rarely apply. When the size of a 2D heterostructure is smaller than a characteristic width $W$ (in the order of 10 nm), strictly speaking, the band alignment is ill-defined. In this case, the SBHs, band offsets, or even the band gap can be altered by changing the system size and interfacial conditions.

## 2. Results and discussion

### 2.1 Graphene/h-BN heterostructure

Inset in Figure 2a shows the supercell of the G/BN heterostructure. Following our recent study [41], eight G/BN interfaces with different misorientation angles are considered. While the $y$ direction is periodic, supercell dimensions of roughly $L$ = 4 nm in the $x$ (junction) direction and $H$ = 1.5 nm in the $z$ (vacuum) direction are initially used to compute the SBHs. Later, these dimensions will be increased to obtain more converged results. The stability of these G/BN heterostructures has been assessed by *ab initio* molecular dynamics simulations. As shown in Figure S1 of Supporting Information (SI), the heterostructures with AC and ZZ (with Clar's reconstruction [41]) interfaces are stable at room temperature.

To obtain band alignment, one needs the average electric potential energy difference across the interface ($eV_{int}$), as a result of interfacial charge transfer from one side of the junction to the other. In the supercell approach, the SBH ($\phi_p$) for *p*-type G/BN is given by [40]:

$$\phi_P = I_S^{bulk} - \Phi_M^{bulk} + eV_{int}, \tag{1}$$

where $I_S^{bulk}$ is the ionization potential of *h*-BN, $\Phi_M^{bulk}$ is the work function of graphene [42]. Alternatively, one may compute $\phi_p$ by locating the respective band edges of the component materials in the local density of states (LDOS) of the supercell in Figure 2a:

$$\phi_p = E_F - E_V^{int}, \tag{2}$$

where $E_F$ is the Fermi level of the system, $E_V^{int}$ is the valence band maximum (VBM) at the center of the *h*-BN region. Equating Eq. (1) with Eq. (2), we obtain

$$eV_{int} = (E_F - E_V^{int}) - (I_S^{bulk} - \Phi_M^{bulk}). \tag{3}$$

When $eV_{int} = 0$, $\phi_p$ follows the Schottky-Mott limit [43, 44], which can be calculated independently by placing graphene and *h*-BN in parallel in a common supercell, as illustrated in the inset in Figure 2b. To ensure better convergence, a large vacuum space with $H = 4.5$ nm is used here. By examining the LDOS in Figure 2b, we determine the Schottky-Mott limit to be $\phi_p = 1.69$ eV.

Even at a smaller $H = 1.5$ nm, one sees a systematic trend in $\phi_p$. For example, Table I summaries $\phi_p$ for eight *p*-type G/BN junctions, where the calculated SBHs fall within a narrow range, $\phi_p$ = 2.10 ± 0.08 eV. In contrast, SBH for 3D junctions depends sensitively on the details of the interfacial structure [39, 45-47].

One can understand this sensitivity by examining the electrostatic models in Figure 3. On the interface, the 2D dipole charge distribution of a 3D junction and the quasi-one-dimensional (1D) inhomogeneous dipole charge distribution of a 2D junction may be approximately replaced by uniform planar and linear paired charge distributions, respectively. For the 3D junction, local charge transfer at the interface leads to formation of a 2D capacitor (Figure 3a) and the amount of the transferred charge determines $eV_{int}$ [39, 40]. Due to the strong screening of the 3D materials, the characteristic width of the junction, $W$, is usually small, often within a couple unit cells (on the order of 1 nm) [48-51]. For the 2D case, however, the capacitor is reduced to a 1D dipole line as schematically shown in Figure 3b, where the field lines are no longer confined to a narrow region at the interface, but leaking into the vacuum that has poor screening effect [52]. This implies that $W$ for a 2D junction will be considerably longer than that for 3D, as we will demonstrate latter.

For the 2D system, one can derive analytically the asymptotic limit for $eV_{int}$ [see SI, details of the dipole line model]. In particular, at a distance $\pm x$ away from the interface, the potential energy difference satisfies

$$eV_{int}(x) = 4k\lambda \ln(1+\frac{4d_{MS}}{2x-d_{MS}}), \tag{4}$$

where $k = 1/(4\pi\varepsilon_0)$, $\lambda$ is the line-averaged charge density, and $d_{MS}$ is an effective distance between charged lines. Since $eV_{int}(x \to \infty) \to 0$, this model suggests that the SBH should approach the Schottky-Mott limit [43, 44].

Care has to be taken, however, in the supercell calculation due to the presence of periodic image charges as schematically shown in Figure 4a. The details of the dipole line model by considering the periodic image charges due to supercell approach are described in SI, in which we show that Eq. (4) should be replaced by

$$eV_{int}(L,H) \approx \frac{8k\lambda\pi}{L/d_{MS}} - \Delta_{HL}(L,H), \qquad (5)$$

where the first term accounts for the effect within the 2D plane of primary concern (including in-plane periodic image charges), and the second term accounts for the effect from all other vertically stacked planes, which reads

$$\Delta_{HL}(L,H) = 8k\lambda \sum_{n=0}^{\infty}\sum_{m=1}^{\infty} \{\ln \frac{[(\frac{1}{4}+n)L/d_{MS} - \frac{1}{2}]^2 + m^2(H/d_{MS})^2}{[(\frac{1}{4}+n)L/d_{MS} + \frac{1}{2}]^2 + m^2(H/d_{MS})^2} + \ln \frac{[(\frac{3}{4}+n)L/d_{MS} + \frac{1}{2}]^2 + m^2(H/d_{MS})^2}{[(\frac{3}{4}+n)L/d_{MS} - \frac{1}{2}]^2 + m^2(H/d_{MS})^2}\} \qquad (6)$$

where $L$ and $H$ are the supercell lengths in the $x$ and $z$ directions, respectively. Figure 4b shows that $eV_{int}$ in the supercell approximation depends on not only $L$ but also $H$. When they are both sufficiently large, $eV_{int}$ always vanishes, so $\phi_p$ approaches the Schottky-Mott limit [cf. Eq. (1)].

One may test Eq. (5) by performing a set of DFT calculations at different $L$ and $H$. For example, Figure 5a shows $eV_{int}(H) = eV_{int}(L_0, H)$ at fixed $L_0$ = 9.84 nm for AC interface and 8.52 nm for ZZ interface, respectively. Figures 5b and 5c show, on the other hand, $eV_{int}(L) = eV_{int}(L, H_0)$ for fixed $H_0$ = 4.5 nm. In all cases, a monotonic decrease is observed. In these plots, the results from the dipole line model in Eq. (5) are also given for comparison. It is clear that the dipole line model reproduces the DFT results reasonably well except for small $L$, at which the model no longer holds due possibly to length-dependent interference of the wave function of graphene [53] and to the deviation of the realistic charge transfer from the simplified uniform dipole line. Similar length-dependent decay behavior was found in 1D heterojunction formed by a semiconducting carbon nanotube and metal contact [54]. A slow decay of $eV_{int}(L,H)$ is in line with experimental observation, where a 10 nm region is required to identify the semi-metallic behavior of graphene and semiconducting behavior of *h*-BN in lateral G/BN junctions [20].

Figures 5b and 5c also suggest that the convergence of $eV_{int}$ can be misorientation-angle dependent. For example, $eV_{int}(L)$ decays more rapidly for the AC case than for the ZZ case. This might be attributed to a larger interfacial charge transfer in the former than the latter (1.314 electrons nm$^{-1}$ for AC and 1.205 electrons nm$^{-1}$ for ZZ on the G/BN interface, respectively, from Bader analysis of the DFT results). In principle, a misorientated interface can be seen as a collection of small segments of the AC and ZZ interfaces with their relative portions determined by the misorientation angle. As such, the interfacial charge transfer on a misorientated interface may be approximated by a linear combination of the amounts of charge transfers at the AC and ZZ interfaces. Known the interfacial charge transfer, one can use Eq. (4) to estimate *W*, which is distance to the interface in the *x* direction when $eV_{int}(L)$ falls within a predetermined small value $\varepsilon$. Figure 6a gives the results for $\varepsilon = 0.05$ eV, showing a ratio between *W*(ZZ) and *W*(AC) to be 11/14=0.786 and an anisotropic dependence of *W* on the misorientation angle.

Compared to the 3D heterostructures whose *W* is typically around 1 nm [48-51], the present results reveal a much longer *W* by up to one order of magnitude (e.g., ~10 nm in G/BN interfaces, ~30 nm in MoS$_2$/WS$_2$ interfaces, see Figure 6). This finding raises a question: what happens if sample dimension *D* is smaller than *W*? In such case, strictly speaking, band alignment is no longer a uniquely-defined quantity, thus SBHs and band offsets should show great sensitivity to sample size and interfacial structure, as well as to interfacial chemistry.

*2.2 MoS$_2$/WS$_2$ heterostructures*

Note that the existence of an interfacial dipole potential is not a property of the metal-semiconductor interfaces alone, but a general behavior of all lateral heterojunctions. Hence, Eq. (5) should also apply to 2D lateral S/S heterostructures where the band offsets are the key parameters for carrier transport. As a demonstration, we consider lateral MoS$_2$/WS$_2$ heterostructures with AC and ZZ interfaces. By analogy, one can expect that the band offsets for a S/S junction follow the Anderson limit [55] also, which is 0.25 eV for MoS$_2$/WS$_2$ according to our PBE calculation.

Figures 5d and 5f show the asymptotic behaviors of $eV_{int}(L,H)$. In contrast to G/BN, however, here the convergence is noticeably slower especially for $H$. This can be attributed to the fact that both MoS$_2$ and WS$_2$ have three atomic layers, rather than a single layer as G/BN. To take this finite thickness into account, Eq. (5) is modified to include a ribbon of thickness $h$ [see the schematic in Figure 4c]. Detailed derivation is given in SI. The model results are given in Figures 5e-5f, showing good agreement with the DFT results. Note that at $L$ = 13 nm, $eV_{int}(L) >$ 0.05 eV. To reach $\varepsilon$ = 0.05 eV, the dipole model predicts $L$ = 21 nm, which would be too large for current DFT calculations. On the other hand, this choice for $\varepsilon$ is arbitrary and as such it should not alter the qualitative picture that the Anderson limit is the correct limit for lateral

MoS$_2$/WS$_2$ heterostructures. Also, lateral MoS$_2$/WS$_2$ junction can be noticeably different for lateral G/BN junction in that the orientation anisotropy in *W* is much smaller (see Figure 6b).

Therefore, in the design of 2D devices using lateral M/S or S/S heterostructures, one can directly use the intrinsic electronic band parameters of each component material without conducting a supercell calculation of the junction [56], as long as the sample size is large enough. At present, experimentally realized samples for *h*-BN, graphene, MoS$_2$, and WS$_2$ typically have a domain size of hundreds to thousands nanometers [14, 17, 19, 26-28], much larger than the characteristic junction width *W*. In such cases, as supported by DFT calculations and guaranteed by classical electrostatic theory, any local changes at the interface will not affect the long-range electrostatic potential alignment of an in-plane 2D heterostructure. In spite of the insensitivity of the band alignment, on the other hand, the transmission function of the 2D heterostructure may still be affected by the interfacial details [57-62].

## 3. Conclusion

To summarize, our combined study of first-principles calculations and analytic model lays the ground for understanding the intrinsic band alignments in broadly-defined 2D lateral heterostructures, which is expected to dictate future experiment and design of 2D electronic/optoelectronic devices. We show the fundamental differences between 2D and 3D junctions, not just by a simple electrostatic argument but by rigorous and extensive first-principles calculations. In particular, when the dimension of a device is considerably larger than the characteristic junction width *W* (which can be one order of magnitude longer than that in 3D), band alignments in the 2D lateral heterostructures should follow the Schottky-Mott (M/S) and Anderson (S/S) limits, respectively, and are insensitive to interfacial details. When such a condition is not satisfied, the band alignment is ill defined, and thus the SBHs or band offsets can be tailored by the component domain size and interfacial conditions.

## 4. Computational Methods

Our calculation employs the density functional theory (DFT) and the projector-augmented wave method (PAW) [63], as implemented in the Vienna *ab initio* simulation package (VASP) [64, 65]. The Perdew–Burke–Ernzerhof (PBE) [66] functional is used to describe the exchange-correlation interactions. The cutoff energy for the plane-wave expansion is 550 eV. The 2D Brillouin zones are sampled by a series of **k** point grids with a constant separations of 0.015 Å$^{-1}$ along the interface direction to ensure the convergence. Supercell of lateral G/BN or $MoS_2$/$WS_2$ junction was constructed by merging two nanoribbons in one plane. The details can be found in our recent paper on in-plane G/BN heterostructures [41].

Generally speaking, there are two types of contributions to the band alignment: (1) the band edges with respect to the electrostatic potential within each component material of the heterostructure; (2) the alignment of the electrostatic potentials across the interface. The first part is sensitive to the functional form used in the calculation and PBE is often insufficient. However, our present conclusion concerns mainly with the second part, namely, the long-range electrostatic potential which depends only on the charge distribution. Since DFT is known to yield rather reasonable charge distribution, the difference between PBE and the hybrid functional like HSE06 and $G_0W_0$ calculation should be zero as the latter uses the DFT charge distribution. More importantly, the electrostatic potential alignment is accurately determined by classical electrostatic theory. Therefore, our current conclusion is expected to be valid regardless computational methods.

ASSOCIATED CONTENT

**Supporting Information**.

Details on the dipole line model. This material is available free of charge via the Internet at http://iopscience.iop.org.

ACKNOWLEDGMENT

Work in China was supported by the National Natural Science Foundation of China (11304191, 11574040), the Fundamental Research Funds for the Central Universities of China (DUT16LAB01), the Supercomputing Center of Dalian University of Technology, and the Natural Science Foundation of Shanxi province (2015021011). WYX was supported by the US-NSF under Award No. 1305293#and SBZ was supported by the US-DOE under Grant No. DESC0002623. The supercomputer time by NERSC under DOE contract No. DE-AC02-05CH11231 and by the CCI at RPI are also acknowledged.

**Table 1.** Supercell lengthes in this work and the corresponding p-type Schottky barrier heights ($\phi_p$) of G/BN.

| Misorientation angle (°) | 0 (= AC) | 9.4 | 13.2 | 21.8 | 32.2 | 38.2 | 42.1 | 60 (= ZZ) |
|---|---|---|---|---|---|---|---|---|
| $L$ (nm) | 3.94 | 3.89 | 4.24 | 3.90 | 4.05 | 3.94 | 3.78 | 3.84 |
| $\phi_p$ (eV) ($H$ = 1.5 nm) | 2.18 | 2.07 | 2.05 | 2.05 | 2.05 | 2.02 | 2.05 | 2.18 |

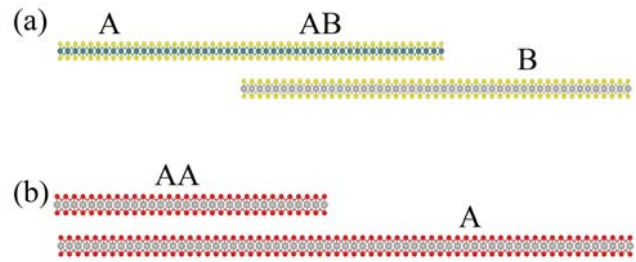

**Figure 1.** Schematic plots for heterostructures with (a) A/AB/B and (b) AA/A structures.

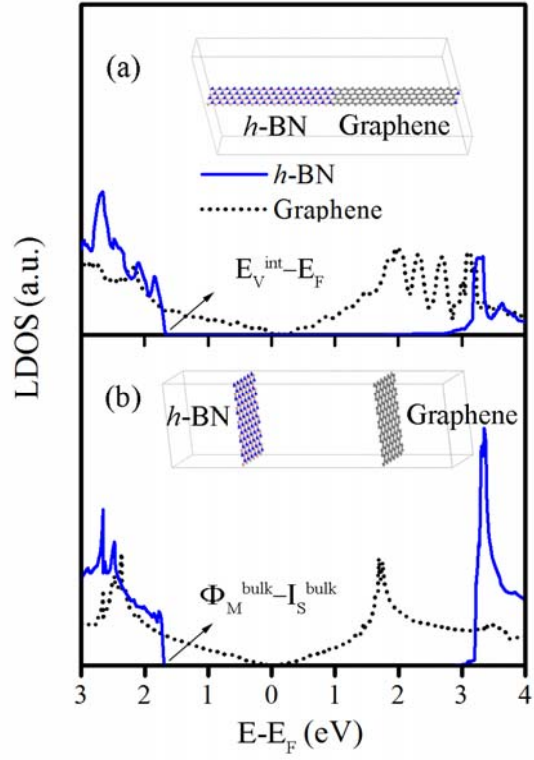

**Figure 2.** LDOS at the centers of the $h$-BN and G regions of (a) lateral G/BN with $L$ = 8.52 nm and $H$ = 4.5 nm and (b) vertically-stacked G and $h$-BN separated by 4.5 nm vacuum. In (a), $|E_F - E_V^{int}|$ = 1.79 eV, whereas in (b), $|\Phi_M^{bulk} - I_s^{bulk}|$ = 1.69 eV, which is the Schottky-Mott limit obtained by using a common vacuum level. Insets are the actual superlattices.

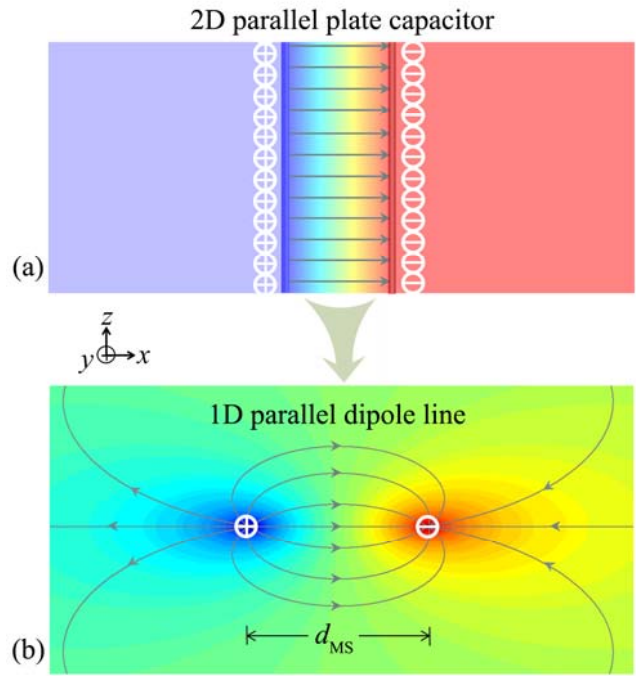

**Figure 3.** (a) 2D capacitor model for a 3D junction and (b) 1D dipole line model for a 2D junction. Potential energy surfaces are shown along with electric field lines. Positive and negative charges are marked.

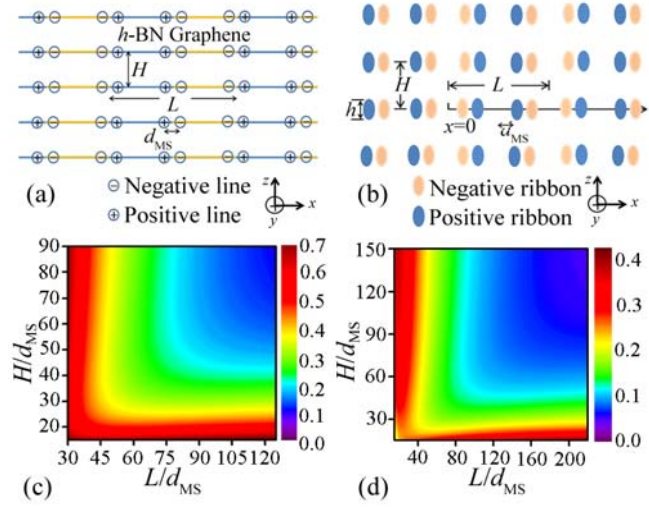

**Figure 4.** Arrangements of (a) charged lines for G/BN and (b) finite-width ribbons for MoS$_2$/WS$_2$ with periodic boundary condition. The corresponding e$V_{int}$ (in unit of eV) as functions of $L$ and $H$ are given in (c) and (d) for G/BN and MoS$_2$/WS$_2$, respectively. Note that in the plots $L$ and $H$ have been rescaled by $d_{MS}$.

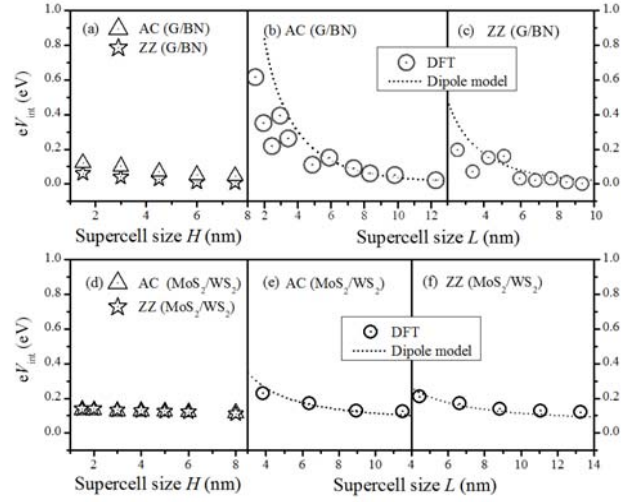

**Figure 5.** Interfacial energy change $eV_{int}$ versus supercell sizes for (a-c) G/BN and (d-f) MoS$_2$/WS$_2$. In (a) $L$ = 8.52 nm for ZZ and 9.84 nm for AC; in (d) $L$ = 11.54 nm for ZZ and 6.38 nm for AC; and in (b), (c), (e) and (f), $H$ = 4.5 nm. Dashed lines in (b) and (c) are the dipole model results for G/BN with $\lambda$ = 0.19 electrons/nm for AC and 0.11 electrons/nm for ZZ, and $d_{MS}$ = 0.05 nm, whereas those in (e) and (f) are the corresponding results for MoS$_2$/WS$_2$ with $\sigma$ = 0.18 electrons/nm$^2$, $d_{MS}$ = 0.1 nm, and $h$ = 0.7 nm. The details of model and definition of the parameters can be found in SI.

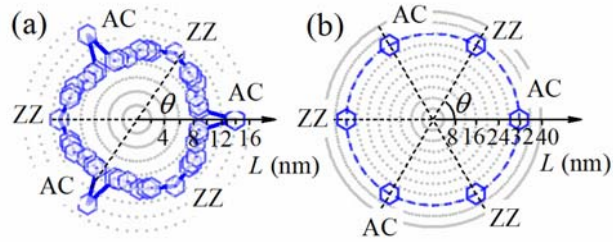

**Figure 6.** Junction width $W$ (for $\varepsilon = 0.05$ eV) calculated using Eq.4, as a function of misorientation angle $\theta$ for (a) G/BN and (b) MoS$_2$/WS$_2$, respectively. The parameters $\lambda$ and $d_{MS}$ come from fitting of the DFT supercell calculations in Figure 5.

# Supporting Information

## Band alignment of two-dimensional lateral heterostructures

Junfeng Zhang[1,2], Weiyu Xie[3], Jijun Zhao[2*], Shengbai Zhang[3*]

[1]*School of Physics and Information Engineering, Shanxi Normal University, Linfen 041004, China*

[2]*Key Laboratory of Materials Modification by Laser, Ion and Electron Beams (Dalian University of Technology), Ministry of Education, Dalian 116024, China*

[3]*Department of Physics, Applied Physics, and Astronomy, Rensselaer Polytechnic Institute, Troy, NY 12180, USA*

* Email: zhaojj@dlut.edu.cn (J. J. Zhao), zhangs9@rpi.edu (S. B. Zhang).

**S1. Details of the dipole line model.**

The electric potential at a position $r$ from a single charge line is

$$V = 2k\lambda \ln \frac{r_0}{r},$$

where $k$ and $\lambda$ are the Coulomb's constant and line charge density, respectively, and $r_0$ is a reference point at which $V = 0$. For a single pair of dipole lines, separated by $d_{MS}$, in Figure 3b, the electrostatic potential difference between points $-x$ and $x$ is given by [1]

$$eV_{int}(x) = 4k\lambda \ln\left(1 + \frac{4d_{MS}}{2x - d_{MS}}\right). \tag{S1}$$

In the DFT calculations, we, however, use the periodic boundary condition, as shown in Figure 4a. Moreover, as shown in Figure S2, Bader analysis for lateral G/BN heterostructures indicated that negative charge always at graphene domain side (see Figure 4a). If we set $(x, z) = (0, 0)$ at the center of graphene, then the electric potential from all negative charge lines would be

$$\phi_N(0,0) = -4k\lambda \sum_{n=0}^{\infty}[\ln \frac{r_0}{\frac{L}{4} - \frac{d_{MS}}{2} + nL} + \ln \frac{r_0}{\frac{3L}{4} + \frac{d_{MS}}{2} + nL}$$
$$+2\sum_{m=1}^{\infty}(\ln \frac{r_0}{\sqrt{(\frac{L}{4} - \frac{d_{MS}}{2} + nL)^2 + (mH)^2}} + \ln \frac{r_0}{\sqrt{(\frac{3L}{4} + \frac{d_{MS}}{2} + nL)^2 + (mH)^2}})],$$

where $L$ is the supercell size and n is the supercell index in the $x$ direction, and $H$ is the supercell size and $m$ is the supercell index in the $z$ direction, respectively. Similarly, the electric potential from all positive charge lines would be

$$\phi_P(0,0) = 4k\lambda \sum_{n=0}^{\infty} [\ln \frac{r_0}{\frac{L}{4} + \frac{d_{MS}}{2} + nL} + \ln \frac{r_0}{\frac{3L}{4} - \frac{d_{MS}}{2} + nL}$$

$$+2\sum_{m=1}^{\infty}(\ln \frac{r_0}{\sqrt{(\frac{L}{4} + \frac{d_{MS}}{2} + nL)^2 + (mH)^2}} + \ln \frac{r_0}{\sqrt{(\frac{3L}{4} - \frac{d_{MS}}{2} + nL)^2 + (mH)^2}})]$$

At the center of graphene $(x, z) = (0, 0)$, the combined electric potential is therefore

$$\phi_G(0,0) = \phi_P(0,0) + \phi_N(0,0)$$

$$= 4k\lambda \sum_{n=0}^{\infty} [\ln \frac{\frac{L}{4} - \frac{d_{MS}}{2} + nL}{\frac{L}{4} + \frac{d_{MS}}{2} + nL} + \ln \frac{\frac{3L}{4} + \frac{d_{MS}}{2} + nL}{\frac{3L}{4} - \frac{d_{MS}}{2} + nL}$$

$$+\sum_{m=1}^{\infty}(\ln \frac{(\frac{L}{4} - \frac{d_{MS}}{2} + nL)^2 + (mH)^2}{(\frac{L}{4} + \frac{d_{MS}}{2} + nL)^2 + (mH)^2} + \ln \frac{(\frac{3L}{4} + \frac{d_{MS}}{2} + nL)^2 + (mH)^2}{(\frac{3L}{4} - \frac{d_{MS}}{2} + nL)^2 + (mH)^2})]$$

The electric potential at the center of $h$-BN can be obtained similarly and it satisfies

$$\phi_{BN}(\text{center}) = -\phi_G(\text{center}).$$

The interfacial potential energy difference is thus

$$eV_{int}(L, H) = \phi_{BN} - \phi_G$$

$$= -8k\lambda \sum_{n=0}^{\infty} [\ln \frac{\frac{L}{4} - \frac{d_{MS}}{2} + nL}{\frac{L}{4} + \frac{d_{MS}}{2} + nL} + \ln \frac{\frac{3L}{4} + \frac{d_{MS}}{2} + nL}{\frac{3L}{4} - \frac{d_{MS}}{2} + nL}$$

$$+\sum_{m=1}^{\infty}(\ln \frac{(\frac{L}{4} - \frac{d_{MS}}{2} + nL)^2 + (mH)^2}{(\frac{L}{4} + \frac{d_{MS}}{2} + nL)^2 + (mH)^2} + \ln \frac{(\frac{3L}{4} + \frac{d_{MS}}{2} + nL)^2 + (mH)^2}{(\frac{3L}{4} - \frac{d_{MS}}{2} + nL)^2 + (mH)^2})]$$

Since $d \ll L$, we can approximate the potential energy difference by

$$eV_{int}(L, H) \approx \frac{8k\lambda\pi}{L/d_{MS}} - \Delta_{HL}(L, H), \tag{S2}$$

where

$$\Delta_{HL}(L,H) = 8k\lambda \sum_{n=0}^{\infty}\sum_{m=1}^{\infty}\{\ln\frac{[(\frac{1}{4}+n)L/d_{MS}-\frac{1}{2})]^2+m^2(H/d_{MS})^2}{[(\frac{1}{4}+n)L/d_{MS}+\frac{1}{2})]^2+m^2(H/d_{MS})^2}+\ln\frac{[(\frac{3}{4}+n)L/d_{MS}+\frac{1}{2})]^2+m^2(H/d_{MS})^2}{[(\frac{3}{4}+n)L/d_{MS}-\frac{1}{2})]^2+m^2(H/d_{MS})^2}\}.$$

For lateral MoS$_2$/WS$_2$ heterostructures, there is a finite layer thickness ($h$) as shown in Figure 3b. Here, the charge lines may be replaced by charge ribbons with a plane charge density $\sigma$. The electric potential at a position $r$ from a single ribbon is thus

$$\phi(x) = \int_{-h/2}^{h/2} 2k\sigma \ln\frac{r_0}{r}dh = \int_{-h/2}^{h/2} 2k\sigma \frac{1}{2}\ln\frac{\frac{h^2}{4}+x_0^2}{\frac{h^2}{4}+x^2}dh \qquad (S3)$$

$$= k\sigma[\int_{-h/2}^{h/2} \ln(h^2+4x_0^2)dh - \int_{-h/2}^{h/2} \ln(h^2+4x^2)dh]$$

Similar to the dipole line model, we have to take into account the effect due to the periodic boundary condition in our numerical calculations.

## S2. Ab initio molecular dynamics simulation of G/BN heterostructures

We investigate the stability of G/BN heterostructures using *ab initio* molecular dynamics (AIMD) simulations. The AIMD simulation was performed with the canonical NVT ensemble using the algorithm by Nosé [2] at 300 K. Each AIMD calculation lasted 3.0 ps, and each MD step for ionic movement was 1.0 fs. Up to 3 ps, there is no noticeable change in both structure and free energy for the systems with armchair (AC) and zigzag (ZZ) interfaces, as demonstrated in Figure S1. Note that the present zigzag interface structure was constructed following Clar's rule [3].

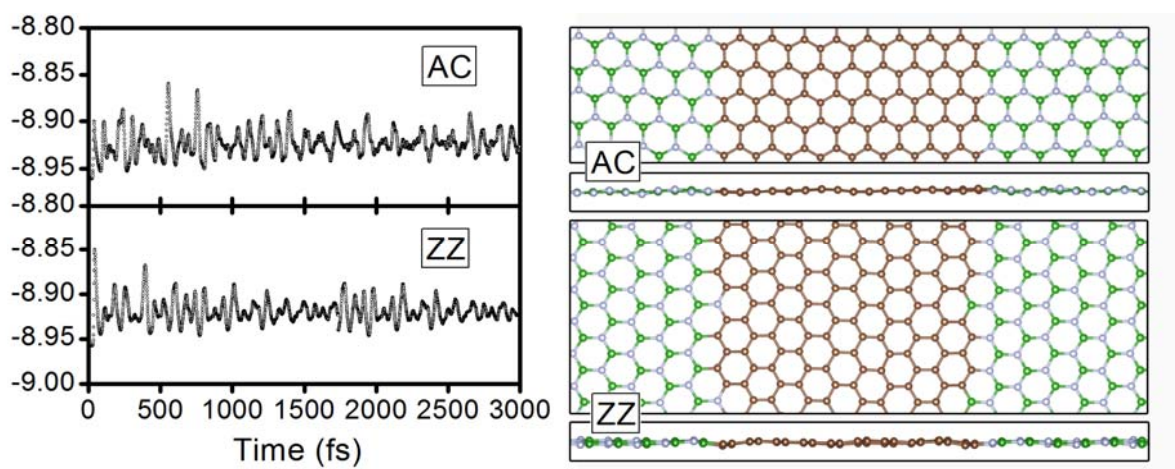

**Figure S1.** (Left) Free energy per atoms for G/BN heterostructures with AC and ZZ interfaces at 300K. (Right) Snapshot structures of G/BN heterostructures with AC and ZZ interfaces (from top view and side view, respectively) from AIMD simulations at 3 ps.

## S3. Bader charge analysis of G/BN heterostructures

We investigate the charge transfer between the graphene and h-BN domains using Bader charge analysis, which indicates 1.314 electrons/nm for AC and 1.205 electrons/nm for ZZ per effective unit length on the G/BN interface. Moreover, as shown in Figure S2, the positive charge is always on the BN side and the negative charge is on the graphene side.

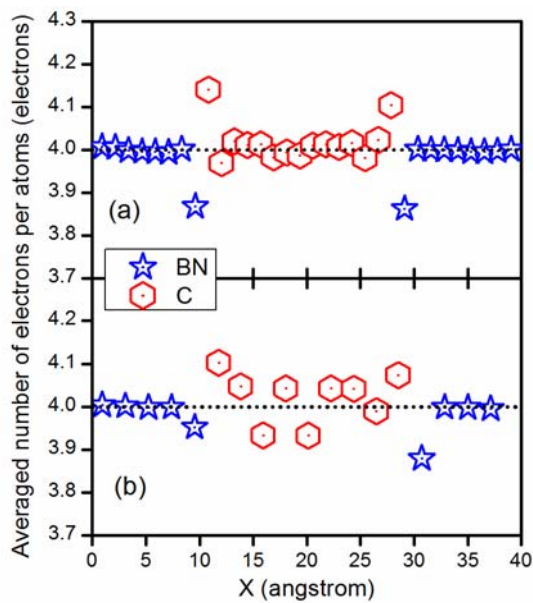

**Figure S2**. Averaged number of electrons per atoms in G/BN heterostructures with AC (a) and cZZ (b) interfaces.